\documentclass[10pt,twocolumn,twoside]{elsart}

\usepackage{graphicx}
\usepackage{amsmath}

\begin{document}

\begin{frontmatter}

\title{Scaling of Anisotropic Flow and Momentum-Space Densities for Light Particles
in Intermediate Energy Heavy Ion Collisions}

\author{T. Z. Yan$^{a,b}$,} \author{ Y. G. Ma$^{a}$ } \ead{
ygma@sinap.ac.cn. Corresponding author.  }
\author{,  X. Z. Cai$^{a}$, J. G. Chen$^{a}$, D. Q. Fang$^{a}$, W. Guo$^{a,b}$,}
\author{   C. W. Ma$^{a,b}$,  E. J. Ma$^{a,b}$, W. Q.
Shen$^{a}$},
\author{ W. D. Tian$^{a}$, K. Wang$^{a,b}$}
\address{$^a$ Shanghai Institute of Applied Physics, Chinese Academy of
Sciences, Shanghai 201800, China\\
$^b$ Graduate School of the Chinese Academy of Sciences, Beijing
100080, China}

\date{\today}

\begin{abstract}

Anisotropic flows ($v_2$ and $v_4$)  of light nuclear clusters are
studied by Isospin-Dependent Quantum Molecular Dynamics model for
the system of $^{86}$Kr + $^{124}$Sn at intermediate energy and
large impact parameters. Number-of-nucleon scaling of the elliptic
flow ($v_2$) are demonstrated for the light fragments up to $A$ =
4, and the ratio of $v_4/v_2^2$ shows a constant value of 1/2. In
addition, the momentum-space densities of different clusters are
also surveyed as functions of  transverse momentum, in-plane
transverse momentum and azimuth angle relative to the reaction
plane. The results can be essentially described by momentum-space
power law. All the above phenomena indicate that there exists a
number-of-nucleon scaling for both anisotropic flow and
momentum-space densities for light clusters, which can be
understood by the coalescence mechanism in nucleonic degree of
freedom for the cluster formation.

\vspace{1pc}
\end{abstract}


\end{frontmatter}

Anisotropic   flow is of interesting subject in  theoretical and
experimental investigations on  nuclear reaction dynamics in
intermediate and high energy heavy-ion collisions
\cite{Olli,Voloshin,Sorge,Danile,Zhang,Shu,Kolb,Zheng,Gale,INDRA}.
Many studies of the dependences of the 1-th and 2-nd anisotropic
flows (the directed flow and elliptic flow, respectively) on beam
energies, mass number or quark number, isospin and impact
parameter revealed much interesting physics about the properties
and origin of the collective flow. In particular, recent
ultra-relativistic Au + Au collision experiments demonstrated the
number of constituent-quark (NCQ) scaling from the transverse
momentum dependent elliptic flow for the different mesons and
baryons at the Relativistic Heavy Ion Collider (RHIC) in
Brookhaven National Laboratory \cite{J.Adams}, it indicates that
the partonic degree of the freedom  plays a dominant role in
formation of the dense matter in the early stage of collisions.
Several theoretical models have been successfully proposed to
interpret the NCQ-scaling of hadrons at RHIC
\cite{Ko,Molnari,Duke,Hwa,Chen}. In these studies, a popular
interpretation is assuming that the mesons and baryons are formed
by the coalescence or recombination of the constituent quarks.
Moving to the intermediate energy heavy ion collisions, the
coalescence mechanism has been also used to explain the formation
of light particles and fragments
\cite{Awes,Mekjian,Sato,Llope,Hagel}. In these studies, however,
most observables which one focused on are the spectra of kinetic
energy or momentum of light particles. While a few studies
investigated the mass dependence of the directed flow
\cite{Huang,Kunde}. Systematic theoretical studies on the flow and
momentum space densities of different fragments in intermediate
energy domain in terms of the coalescence mechanism are still
rare. In this context,
 we survey, for the first time  to our knowledge,
  the nucleon number dependence of  the  anisotropic
flows $v_2$ and $v_4$ in the intermediate energy heavy ion
collisions with the coalescence scenario. Moreover, the power law
behaviors of light fragment production in momentum space are also
explored in our model simulation. Note that this power law
behaviors have been experimentally demonstrated in violent
collisions at the beam energies between $0.1$ A and $15$ A GeV
\cite{H.H.Gutbrod}. In this Letter we use Isospin-Dependent
Molecular Dynamics (IDQMD) model  to simulate $^{86}$Kr +
$^{124}$Sn at 25 MeV/nucleon and larger impact parameters (b = 7 -
10 fm), and investigate the outcome of nucleonic coalescence
mechanism on the anisotropic flows and momentum space densities of
light particle production in intermediate energy heay ion
collisions.

Anisotropic flow is defined as the different $n$-th harmonic
coefficient $v_n$ of an azimuthal Fourier expansion of the
particle invariant distribution \cite{Voloshin}
\begin{equation}
\frac{dN}{d\phi} \propto {1 + 2\sum_{n=1}^\infty v_n cos(n\phi) },
\end{equation}
where $\phi$ is the azimuthal angle between the transverse
momentum of the particle and the reaction plane. Note that in the
coordinate system the $z$-axis along the beam axis, and the impact
parameter axis is labelled as $x$-axis.  The first harmonic
coefficient $v_1$ represents the directed flow,
$v_1 = \langle cos\phi \rangle = \langle \frac{p_x}{p_t} \rangle$,
where $p_t = \sqrt{p_x^2+p_y^2}$ is transverse momentum. $v_2$
represents the elliptic flow which characterizes the eccentricity
of the particle distribution in momentum space,
\begin{equation}
v_2 = \langle cos(2\phi) \rangle = \langle
\frac{p^2_x-p^2_y}{p^2_t} \rangle,
\end{equation}
and $v_4$ represents the 4-th momentum anisotropy,
\begin{equation}
v_{4} =\left\langle \frac{p_{x}^{4}-6p_{x}^{2}p_{y}^{2}+p_{y}^{4}}{%
p_{t}^{4}}\right\rangle .
\label{v4}
\end{equation}

The model we are using in the present work is based on the Quantum
Molecular Dynamics (QMD) approach which is an n-body theory to
describe heavy ion reactions from intermediate energy to 2 A GeV.
It includes several important parts: the initialization of the
target and the projectile nucleons, the propagation of nucleons in
the effective potential, the collisions between the nucleons, the
Pauli blocking effect and the numerical tests. A general review
about QMD model can be found in \cite{Aichelin}. The IDQMD model
is based on QMD model affiliating the isospin factors, which
includes the mean field, two-body nucleon-nucleon (NN) collisions
and Pauli blocking \cite{Ma3,Liu,Wei,Ma2,Ma-hbt}.

In the QMD model each nucleon is represented by a Gaussian wave
packet with a width $\sqrt{L}$ (here $L$ = 2.16 ${\rm fm}^2$)
centered around the mean position $\vec{r_i}(t)$ and the mean
momentum $\vec{p_i}(t)$,
\begin{equation}
\psi_i(\vec{r},t) = \frac{1}{{(2\pi L)}^{3/4}}
exp[-\frac{{(\vec{r}- \vec{r_i}(t))}^2}{4L}] \nonumber
\end{equation}
\begin{equation}
exp[-\frac{i\vec{r} \cdot \vec{p_i}(t)}{\hbar}].
\end{equation}

The nucleons interact via nuclear mean field  and nucleon-nucleon
collision. The nuclear mean field can be parameterized by
\begin{equation}
U(\rho,\tau_{z}) = \alpha(\frac{\rho}{\rho_{0}}) +
\beta(\frac{\rho}{\rho_{0}})^{\gamma} +
\frac{1}{2}(1-\tau_{z})V_{c} \nonumber
\end{equation}
\begin{equation}
+ C_{sym} \frac{(\rho_{n} - \rho_{p})}{\rho_{0}}\tau_{z} + U^{Yuk}
\end{equation}
with $\rho_{0}$ the normal nuclear matter density (here, 0.16
$fm^{-3}$ is used). $\rho$, $\rho_{n}$ and $\rho_{p}$ are the
total, neutron and proton densities, respectively. $\tau_{z}$ is
$z$th component of the isospin degree of freedom, which equals 1
or -1 for neutrons or protons, respectively. The coefficients
$\alpha$, $\beta$ and $\gamma$ are parameters for nuclear equation
of state (EOS). $C_{sym}$ is the symmetry energy strength due to
the difference of neutron and proton. In the present work, we take
$\alpha$ = -124 MeV, $\beta$ = 70.5 MeV and $\gamma$ = 2.0 which
corresponds to the so-called hard EOS with an incompressibility of
$K$ = 380 MeV and $C_{sym}$ = 32 MeV \cite{Aichelin}. $V_{c}$ is
the Coulomb potential and $U^{Yuk}$ is Yukawa (surface) potential
which has the following form:

\begin{equation}
U^{Yuk} =
\frac{V_{y}}{2m}\sum_{{i}\neq{j}}\frac{1}{r_{ij}}exp(Lm^{2})\nonumber
\end{equation}
\begin{equation}
[exp(-mr_{ij})erf(\sqrt{L}m -r_{ij}/\sqrt{4L}) \nonumber
\end{equation}
\begin{equation}
- exp(mr_{ij})erf(\sqrt{L}m+r_{ij}/\sqrt{4L})]
\end{equation}
with $V_{y}$ = 0.0074 GeV, $m$ = 1.25 $fm^{-1}$ and $L$ = 2.16 $
fm^{2}$. The relative distance $r_{ij} = |\vec{r_i}-\vec{r_j}|$.
Experimental in-medium NN cross section parametrization which is
energy and isospin dependent is used in this work.

The Pauli blocking effect in IDQMD model is treated  separately
for  the neutron and the proton: whenever a collision occurs,  we
assume that each nucleon occupies a six-dimensional sphere with a
volume of $\hbar^{3}$/2 in the phase space (considering the spin
degree of freedom), and then calculate the phase volume, $V$, of
the scattered nucleons being occupied by the rest nucleons with
the same isospin as that of the scattered ones. We then compare
2$V$/$\hbar^{3}$ with a random number and decide whether the
collision is blocked or not.

In the QMD model, the initial momentum  of nucleons is generated
by means of the local Fermi gas approximation. The local Fermi
momentum is given by:
 \begin{equation}
P^{i}_{F}(\vec r) = \hbar(3\pi^{2}\rho_{i}(\vec r))^{\frac{1}{3}},
(i= n,p).
\end{equation}
In the model, the radial density can be written as:
\begin{equation}
\rho(r) = \sum_{i}\frac{1}{(2\pi L)^{3/2}} exp(-\frac{r^{2} +
r_{i}^{2}}{2L})\frac{L}{2rr_{i}} \nonumber
\end{equation}

\begin{equation}
\times[exp(\frac{rr_{i}}{L}) - exp(-\frac{rr_{i}}{L})].
\end{equation}

The time evolution of the colliding system is given by the
generalized variational principal. Since the QMD can naturally
describe the fluctuation and correlation, we can study the nuclear
clusters in the model \cite{Aichelin,Ma3,Liu,Wei,Ma2,Ma-hbt}.
 In QMD model, nuclear clusters are usually recognized  by a simple
coalescence model: i.e. nucleons are considered to be part of a
cluster if in the end at least one other nucleon is closer than
$r_{min} \leq 3.5$ fm in coordinate space and $p_{min} \leq 300$
MeV/c  in momentum space  \cite{Aichelin}. This
mechanism has been extensively applied in transport theory for the
cluster formation.

\begin{figure}
\vspace{-0.2truein}
\includegraphics[scale=0.35]{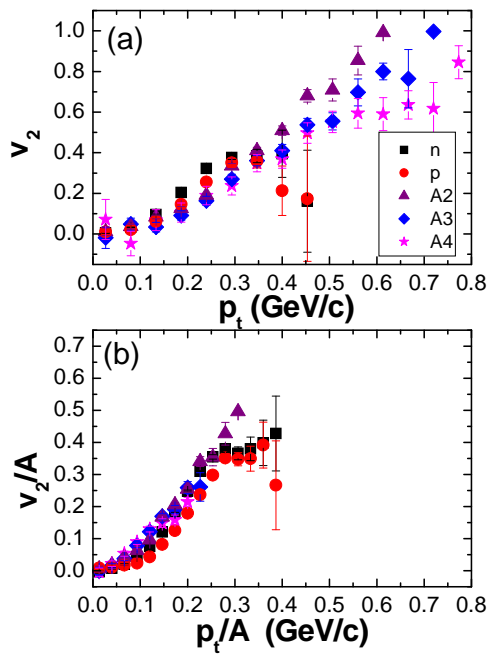}
\vspace{0.2truein} \caption{\footnotesize (a) Elliptic  flow as a
function of transverse momentum ($p_t$). Squares represent for
neutrons, circles for protons, triangles for fragments of $A$ = 2,
diamonds for $A$ = 3 and stars for $A$ = 4; (b) Elliptic flow per
nucleon as a function of transverse momentum per nucleon.  The
symbols are the same as (a). } \label{Fig-v2-pt}
\end{figure}

Now we move to the calculations. For an example, we simulated
$^{86}$Kr + $^{124}$Sn at 25 MeV/nucleon and impact parameter of 7
- 10 fm. 50,000 events have been accumulated. The systems tend to
freeze-out around 120 fm/c. In this paper, we extract the
following physics results at 200 fm/c.  The upper panel of
Fig.~\ref{Fig-v2-pt} shows transverse momentum dependence of
elliptic flows for mid-rapidity light fragments. The range of
transverse momentum for different fragments is different according
to their masses. From the figure, it shows that the elliptic flow
is positive and it increases with the increasing $p_t$. It
reflects that the light clusters are preferentially emitted within
the reaction plane, and  particles with higher transverse momentum
tend to be strongly  emitted within in-plane , i.e. stronger
positive elliptic flow. In comparison to the elliptic flow at RHIC
energies, the apparent behavior of elliptic flow versus $p_t$
looks similar, but the mechanism is obviously different. In
intermediate energy domain, collective rotation is one of the main
mechanisms to induce the positive elliptic flow
\cite{Ritter,Peter,Shen,yg-prc,Lacey,He}. In this case, the
elliptic flow is mainly driven by the attractive mean field.
However, the strong pressure which is built in early initial
geometrical almond-type anisotropy due to the overlap zone between
both colliding nuclei in coordinate space will rapidly transforms
into the azimuthal anisotropy in momentum space at RHIC energies
\cite{J.Adams}. In other words, the elliptic flow is mainly driven
by the stronger outward pressure. The lower panel displays the
elliptic flow per nucleon as a function of transverse momentum per
nucleon, and it looks that there exists the number of nucleon
scaling when $p_t/A < 0.25 $ GeV/$c$. This behavior is apparently
similar to the number of constituent quarks scaling of elliptic
flow versus transverse momentum per constituent quark ($p_t/n$)
for mesons and baryons which was observed at RHIC \cite{J.Adams}.

The RHIC experimental data demonstrated a scaling relationship
between 2-nd flow ($v_2$) and n-th flow ($v_n$), namely
$v_{n}(p_{t})\sim v_{2}^{n/2}(p_{t})$ \cite{STAR03}. It has been
shown \cite{Kolb2,ChenLW} that such scaling relation follows from
a naive quark coalescence model \cite{Molnari} that only allows
quarks with equal momentum to form a hadron. Denoting the meson
anisotropic flows by $v_{n,M}(p_t)$ and baryon anisotropic flows
by $v_{n,B}(p_t)$ , Kolb et al. found that $v_{4,M}(p_{t}) =
(1/4)v_{2,M}^{2}(p_{t})$ for mesons and $v_{4,B}(p_{t}) =
(1/3)v_{2,B}^{2}(p_{t})$ for baryons if quarks have no
higher-order anisotropic flows. Since mesons dominate the yield of
charged particles in RHIC experimental data, the smaller scaling
factor of $1/4$ than the empirical value of about $1$ indicates
that higher-order quark anisotropic flows cannot be neglected.
Including the latter contribution, one can show that
\begin{equation}
\frac{v_{4,M}}{v_{2,M}^{2}} \approx \frac{1}{4}+\frac{1}{2}%
\frac{v_{4,q}}{v_{2,q}^{2}},\allowbreak \label{v4Mscal}
\end{equation}
and
\begin{equation}
\frac{v_{4,B}}{ v_{2,B}^2} \approx \frac{1}{3} + \frac{1}{3}
\frac{v_{4,q}}{v_{2,q}^2}\,,
\end{equation}
where $v_{n,q}$ denotes the quark anisotropic flows. The meson
anisotropic flows thus satisfy the scaling relations if the quark
anisotropic flows also satisfy such relations. However, this ratio
is experimentally determined to be 1.2 \cite{star_workshop}, which
means that the fourth-harmonic flow of quarks $v_4^q$ must be
greater than zero. One can go one step further and assume that the
observed scaling of the hadronic $v_2$ actually results from a
similar scaling occurring at the partonic level. In this case
$v_4^q = (v_2^q)^2$ and the hadronic ratio $v_4/v_2^2$ then equals
$1/4 + 1/2 = 3/4$ for mesons and $1/3+1/3=2/3$ for baryons,
respectively. Again, since this value is measured to be 1.2, even
the partonic $v_4^q$ must be greater than simple scaling and quark
coalescence models predict.

\begin{figure}
\vspace{-0.2truein}
\includegraphics[scale=0.35]{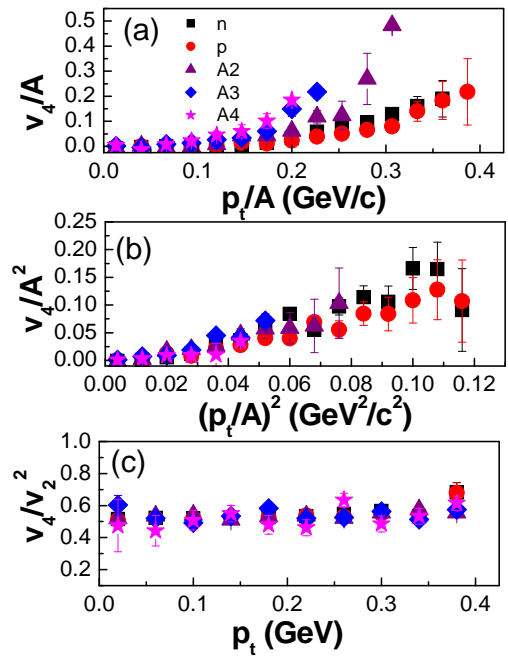}
\vspace{0.2truein} \caption{\footnotesize (a) $v_4/A$ as a
function of $p_t/A$ for different particles, namely, neutrons
(squares), protons (circles), fragments of $A$ = 2 (triangles),
$A$ = 3 (diamonds) and $A$ = 4 (stars). (b) $v_4/A^2$ as a
function of $(p_t/A)^2$. (c) the ratios of $v_4/v_2^2$ for
different particles vs  $p_t$.
 }\label{Fig-v4}
\end{figure}

Recognizing the above behaviors of the flows at RHIC energies, we
would like to know what the higher order momentum anisotropy in
the intermediate energy is. So far, there is neither experimental
data nor theoretical investigation for the higher order flow, such
as $v_4$, in this energy domain. In the present work, we explore
the behavior of $v_4$ in the model calculation. Fig.~\ref{Fig-v4}
shows the feature of $v_4$. Similar to the relationship of $v_2/A$
versus $p_t/A$, we plot $v_4/A$ as a function of $p_t/A$. The
divergence of the different curves between different particles in
Fig.~\ref{Fig-v4}(a) indicates no simple scaling of nucleon number
for 4-th momentum anisotropy. However, if we plot $v_4/A^2$ versus
$(p_t/A)^2$, it looks that the points of different particles
nearly merge together and it means a certain of scaling holds
between two variables. Due to a nearly constant value of
$v_4/v_2^2$ in the studied $p_t$ range (see Fig.~\ref{Fig-v4}(c))
together with the number-of-nucleon scaling behavior of $v_2/A$ vs
$p_t/A$, $v_4/A^2$ should scale with $(p_t/A)^2$, as shown in
Fig.~\ref{Fig-v4}(b). If we assume the scaling laws of mesons
(Eq.(9)) and baryons (Eq.(10)) are also valid for A = 2 and 3
nuclear clusters, respectively, then $v_4/v_2^2$ for A = 2 and 3
clusters indeed give the same value of 1/2 as nucleons, as shown
in Fig.~\ref{Fig-v4}(c). Coincidentally the predicted value of the
ratio of $v_4/v_2^2$ for hadrons is also 1/2 if the matter
produced in ultra-relativistic heavy ion collisions reaches to
thermal equilibrium and its subsequent evolution follows the laws
of ideal fluid dynamics \cite{Bro}. It is interesting to note the
same ratio was predicted in two different models at very different
energies, which is of course worth to be further investigated in
near future. Overall speaking, we learn  that $v_4/v_2^2$ is
approximately 1/2 in nucleonic level coalescence mechanism, which
is different from 3/4 for mesons or 2/3 for baryons in partonic
level coalescence mechanism.

Measurements of inclusive single-particle spectra from heavy-ion
collisions indicate a simple empirical pattern of light fragment
production: the observed invariant momentum-space density
${\rho}_A$ for fragments with mass number $A$ closely follows the
$A$-th power of the observed proton density ${\rho}^A_1$. It was
shown that this power law behavior is  valid for spectra of
participant fragments up to $A$ = 14 with projectiles ranging from
protons to Au at a variety of beam energies between 0.1 A and 15 A
GeV \cite{H.H.Gutbrod}. However, there are rare experimental data
and calculations to test the power law at intermediate energies.
In the present work, we test the momentum-space power law up to
$A$ = 4  by IDQMD.

\begin{figure}
\vspace{-.3truein}
\includegraphics[scale=0.38]{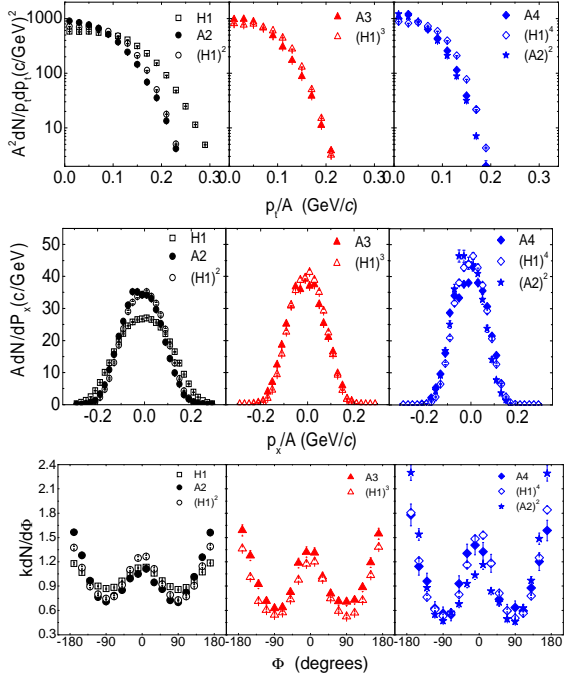}
\caption{\footnotesize Upper panels: transverse momentum-space
density for light fragments; Middle panels: in-plane transverse
momentum space density; Bottom panels: azimuthal distributions
relative to the reaction
 plane. The open squares represent the density for protons, the solid
circles for the fragments with $A$ = 2,
 the open circles for the proton density squared normalized to the same area as
 the  proton density, the solid triangles for $A$ = 3, the open
 triangles for protons to the power of 3, the solid diamonds for
 $A$ = 4, the open diamonds for protons to the power of 4, and the stars
 for the $A$ = 2 density squared. }  \label{Fig-spectra}
\end{figure}

The upper panels of Fig.~\ref{Fig-spectra} show the transverse
momentum space densities $\rho = A^2 dN/p_t dp_t$. The results
also show a level of adherence to power law behavior at high
$p_t/A$, which is similar to what have been previously reported
for single-particle-inclusive measurements \cite{S.Wang}, and
reflects the persistence of momentum-space coalescence behavior
for intermediate energy $^{86}$Kr + $^{124}$Sn collisions. The
middle panels of Fig.~\ref{Fig-spectra} depict the density in the
in-plane transverse momentum space ($p_x/A$). The densities are
also normalized to the same area as the proton density. Wang ${\it
et\ al.}$  first introduced the momentum-space power law to
$\rho(p_x)$ \cite{S.Wang}. Again, our results show a level of
adherence to power law behavior. The bottom panels of
Fig.~\ref{Fig-spectra} depicts the azimuthal ($\phi$) distribution
of fragments relative to the reaction plane. The factor $k$ is
chosen so that the mean values of $dN/d\phi$ are fixed to be 1.
Essentially the power-law behavior remains for the azimuthal
distributions of these light fragments. From all the above
distributions, nucleon coalescence mechanism keeps valid in the
momentum space densities.

In summary, we applied IDQMD model to investigate the behavior of
anisotropic flows, namely $v_2$ and $v_4$,  versus transverse
momentum for the light fragments from 25 MeV/nucleon $^{86}$Kr +
$^{124}$Sn collisions at 7-10 fm of impact parameters. Both $v_2$
and $v_4$ generally show positive values and increase with $p_t$.
By the number-of-nucleon scaling, the curves of elliptic flow for
different fragments approximately collapse on the similar curve,
which means that there exists an elliptic flow scaling on the
nucleon number for light fragments. This phenomenon is similar to
the NCQ scaling of elliptic flows of hadrons at RHIC energies
where elliptic flow can be developed in the early partonic stage
of collisions, but here it is a nucleonic level in intermediate
energy collisions. For 4-th momentum anisotropy $v_4$, it seems to
be scaled by $v_2^2$ and $v_4/v_2^2 \sim 0.5$. It will be of very
interesting if one can measure this ratio in intermediate energy
HIC. In addition, we also investigated momentum-space densities of
light fragments as functions of fragment transverse momentum
$p_t$, in-plane transverse momentum $p_x$ and the azimuth angle
relative to the reaction plane. All these observables are well
described by the momentum-space power law which has been well
experimentally observed at high energies. All the above phenomena
can be seen as an outcome of the nucleonic coalescence which
results that the number-of-nucleon scaling of both flow and
momentum space density in intermediate energy heavy ion collision.
 It may indicate that nucleonic matter may be one of the middle transit
stage before chemical freeze-out takes place.

{\bf Acknowledgements}: This work was supported in part by the
Shanghai Development Foundation for Science and Technology under
Grant Numbers 05XD14021, the  National Natural Science Foundation
of China  under Grant No 10328259, 10135030, and 10535010.

{}

\end{document}